\documentclass[10pt]{article}

\usepackage[english]{babel}
\usepackage{amstext}
\usepackage{amsmath,amssymb,amsthm,bm, float}
\usepackage{graphicx,psfrag,epsf}
\usepackage{enumerate}
\usepackage{color, soul}
\usepackage[colorinlistoftodos]{todonotes}
\usepackage{natbib}
\usepackage{url}
\usepackage{comment}

\usepackage{enumitem}
\usepackage{ulem}
\usepackage{multirow}
\usepackage{setspace}
\usepackage{titlesec}
\titlespacing*{\section}{0pt}{0.05\baselineskip}{0.1\baselineskip}

\newcommand{\upperRomannumeral}[1]{\uppercase\expandafter{\romannumeral#1}}

\newcommand{\resethlcolor}{\sethlcolor{yellow}}
\resethlcolor

\numberwithin{equation}{section}

\def\spacingset#1{\renewcommand{\baselinestretch}%
{#1}\small\normalsize} \spacingset{1.45}
\title{\bf{Conducting Highly Principled Data Science: \\ A Statistician's Job and Joy}\footnote{This article is prepared  for the special issue on ``The Role of Statistics in the Era of Big Data" organized by \textit{Statistics and Probability Letters}. I thank the editor Laura M. Sangalli for inviting me, my Astrostatistics collaborators for making my adventure possible, Joe Blitzstein, Yang Chen, Radu Craiu, Francesca Dominici, Vinay Kashyap, Todd Kuffner, Bharmar Mukherjee, Aneta Siemiginowska, Lei Sun and a reviewer for comments and encouragements, Steve Finch for proofreading, and the US National Science Foundation for partial financial support.}}

\author{Xiao-Li Meng \\ \small Department of Statistics, Harvard University, Cambridge, MA 02138}

\date{\textit{This article is dedicated to the 20th Birthday of CHASC\footnote{California Harvard Astro-Statistics Collaboration, established in 1997 by statistician David van Dyk and astrophysicists Alanna Connors (Wellesley College), Vinay Kashyap and Aneta Siemiginowska (Harvard-Smithsonian Center for Astrophysics). I helped to lead the statistical team on the Harvard side after David moved to the University of California at Irvine in 2003, and subsequently to Imperial College of London in 2011, which brought CHASC to the international arena. Alanna was a driving force of CHASC's education mission and outreach effort, helping statisticians understand science and scientists understand statistics. She devoted herself to such causes to the very end of her life. She wrote on January 29, 2013, ``My cancer is not responding to any treatment, so I am going into Hospice (at home) today.  I am very tired, so I may not be ab[l]e to participate much. I'll try skyping in tomorrow. With many thanks for everything".  She passed away on February 2, 2013, after more than a decade fighting  with breast cancer.}  \\  and to the memory of Alanna Connors, a founding astrophysicist of CHASC}}
\usepackage{varioref}

\begin{document}
\vskip -1in
\maketitle\ 
\vskip -.5in
\noindent{\textbf{Abstract}}\quad  Highly Principled Data Science insists on methodologies that are: (1) scientifically justified; (2) statistically principled; and (3) computationally efficient. An astrostatistics collaboration, together with some reminiscences, illustrates the increased roles statisticians can and should play to ensure this trio, and to advance the science of data along the way.


\section{Proactive co-investigators/partners, not passive consultants}\label{sec:cons}

On March 26, 2015, I received the following email from an organizer of the 10th International Astronomical Consortium for High-Energy Calibration (IACHEC) meeting, which contains the following question:

``\textit{Systematic errors in comparing effective areas}: {\sl Speaking hypothetically, if we label the instruments by numbers $i=1, \ldots,  N$ and each has an attribute $A$ that is used to measure the same $j=1, \ldots, M$ astrophysical sources, with intrinsic attribute $F_j$ where $C_{ij} = A_i F_j$ are the instrumental measurements, then the question is: `Is there a way to decide how (or whether) to change $A_i$ when the values $C_{ij}/A_i$ do not agree with $F_j$ to within their statistical uncertainties $s_i$. In other words, each instrument provides an estimator $f_j$ of $F_j$ with statistical uncertainty $s_j$ but $|f_j - F_j|/s_j$ is often large, not distributed as a Gaussian with unit variance $\cdots$. How to estimate the systematic error on the $A_i$?}"

Being a member of CHASC, I was invited to give a general statistical tutorial on April 20, 2015, at the IACHEC meeting in Beijing. The email came just about the time I was trying to settle on my tutorial topics. Frankly, up to that point, I had not thought very hard about how to tailor my tutorial towards \textit{the} problem that IACHEC cares about, and indeed its reason of existence, that is, building \textit{concordance} among astronomical instruments operated by different teams. 
Knowing that I was new to this meeting, the same organizer wrote to me a few days earlier, which highlighted this goal:  
``{\sl A few words on the IACHEC: it is a gathering of astronomers involved in the calibration of X-ray instrumentation of past, present (operational) and future missions. Our main goal is improving the mutual agreement between measurements yielded by different instruments to increase the fidelity of the science extracted by high-energy astrophysical data. An important part of this work is the collective setting of standards in, e.g., X-ray data analysis, that may constitute a reference for the whole astronomical community."}

Seeing some specifics, I realized that I could contribute more than giving a tutorial. Minimally I could introduce the concept and calculation of shrinkage estimation, and demonstrate why one needs to avoid the notoriously (to statisticians) unstable ratio estimators, which apparently were what the IACHEC community was using. However, my decade-long involvement with CHASC projects taught me that any time I see a problem with a seemingly obvious solution, I should double check with my scientific collaborators if they have simplified the real messy world, mostly for the benefit of statisticians like me.  Therefore, to be sure that I was not being overly confident, I wrote back to confirm my understanding: {\sl ``
(1) $C_{ij}$ are observations, and it is safe to assume that conditioning on the *true* $A_i$ and $F_j$,   $C_{ij}$'s are independent of each other.\ (2) Both the true $A_i$ and $F_j$ are unknown, but you have some estimates $a_i$ for $A_i$ and $f_j$ for $F_j$ based on some other experiments or theoretical models, and it would be safe to assume that $(f_j, a_i, C_{ij})$ are all independent conditioning on the true (but unknown) values of $A_i$'s and $F_j$'s.\ (3) The question is that,  given the values of $C_{ij}$ and $a_i$ and $f_j$, what are the best estimators of $A_i$ for all $i$ (better than just using $a_i$ for $A_i$)?''}

Vinay (who was cc'ed on my emails), a member of both CHASC and IACHEC, confirmed my suspicion that the problem is harder than it appears to be:
{\sl 
``The goal of IACHEC was to make these measurements of $C_{ij}$ (counts from
source $j$ observed with instrument $i$), and given a knowledge of source
spectrum $f_j$ (often incomplete, but usually known to better than a few
percent), to adjust the instrument response $a_i$ so that all analyses
produce consistent results.  This has been surprisingly difficult to
achieve.  Part of the problem is that $C, a,$ and $f$ are all functions of
energy, and the overlap between the different instruments is not 100\%,
and some instruments are more reliable at some energies compared to others."}

These few email exchanges turned out to be the beginning of hundreds (and counting) of communications---emails, skypes, in-person meetings, workshop exchanges, etc.---in the past two years among a group of astrophysicists and statisticians. This type of exchanges, in terms of both their frequencies and nitty-gritty nature, should come as no surprise to anyone who has engaged in serious interdisciplinary collaborations on challenging problems. \textit{Challenging problems are unsolvable in a few consultation sessions.}  This almost tautological statement lies at the heart of how we statisticians can increase our direct impact on advancing science through data, concurrent with advancing the science of data. 

Ages ago,  I served for three years as the Director of the Consulting Program shortly after I joined The University of Chicago as an assistant professor. I encouraged all students, when they met with their clients, to ask as many questions as possible about the data collection process, emphasizing that nothing is more important than the data quality. Whereas that was the right emphasis, something I would stress even more in this age of big-messy data, I had no experience myself about effective communication with those who were seeking statistical help.  Inevitably, some clients felt that we were overly critical but not very helpful, to the extent  that one of them told us that ``I am here for consultation, not for insultation.''

As I grew professionally, I came to realize, albeit gradually, that the issue went deeper than communication skills. Being overly critical but not constructive is a telling sign of lacking  the feeling of ownership or accountability, neither of which helps to entice the consultants  to invest time or energy as they would for solving their own problems. Nor would the clients feel the urge to inform the consultants about their investigation processes. Indeed, a sizable number of clients to the consulting program then wanted quick answers to questions such as ``What's the \textit{p}-value for this test that a reviewer asked me to perform?". Historically, such attitudes towards statistical analysis had led to decisions to avoid setting up such a program \citep[e.g.,][641-642]{chan2001interview}.

Both the scientific and statistical communities have come a long way since then in seeing the need of working together, not as consultants and clients, but as genuine partners and co-investigators in scientific investigations. To make this partnership truly effective, and mutually beneficial, will require investing time and energy on both sides to understand each other's language, perspectives, and modus operandi. 
For statisticians, to listen and ask critical---but constructive---questions from the very beginning is a crucial first step towards a fruitful collaboration. 
The IACHEC concordance project reminds me once more of the job, and joy, of a statistician in this partnership. It also helped me to crystallize the meaning of conducting highly-principled data science, as I shall elaborate below. But a disclaimer before proceeding:  the opinions expressed below and my choices of the expressions are neither (entirely) new nor (completely) final, and they are inherently idiosyncratic as individual opinions always are. Disagreements are greatly encouraged, as a part of our collective brainstorming about how we can simultaneously broaden our horizons, being a pillar of data science, and deepen our foundations, to earn and ensure our fundamental roles in scientific inquires and discoveries.  

\section{Scientifically justified, not merely motivated}

Years ago, a Chicago colleague told me a story that must sound ridiculous now. Sometime in 1960s, a colleague at his previous institution wrote a grant proposal to a national defense agency, which started with ``Let $X_1, \ldots, X_n$ be an i.i.d. sample, where $X$, for example, is the size of a tank."    
And that's the end of the military involvement. Whether that was designed to test the agency's review capacity beyond the first sentence, or to provide a historical testimony to the deep division between theory and applications,  we surely have passed the era of focusing on justifying the practical relevance of statistics (though, unfortunately, I cannot add ``or statisticians" yet). An obvious benefit for us statisticians to take on a co-investigator role in any collaborative project is that we no longer need to defend our choices of topics with a battalion of tanks.

Of course, there is no free lunch---we face a much more demanding task: help to ensure the process and product of our investigation are \textit{scientifically justified}. Early on,  I naturally regarded \textit{that} as the responsibility of my scientific collaborators, and my job was to translate their goals and knowledge into statistical models and then help on the analyses. I can never have their scientific insights, just as they perhaps would never feel comfortable to question the validity of my statistical methods or results. But gradually I realized that if translation and analysis is all we do, then we have not really gone beyond a consultative mindset. More importantly, as much as this may sound to be a professional self-promotion, we statisticians can and should help to also ensure and enhance scientific justifications.

This is because virtually all scientific investigations are about separating signals from noises, figuratively and literally, and about updating previous understanding and consensus in the presence of new information or ideas or both. But these are the very processes concerning which we statisticians are trained to think deeply, broadly, and rigorously. In a grand scheme of things, the only difference between a domain expert and a statistician is that the former engages in these processes with the goal to advance a substantive field, whereas for the latter the goal and the field is to advance these processes themselves, with or without reference to applications in specific substantive fields.  We statisticians therefore are uniquely equipped to be investigative partners with our scientific collaborators, not merely for the data analysis process. We can and should play increasingly more proactive roles in problem formulation (e.g., define what should constitute as signals and what should be regarded as noises), data collection (e.g., promote efficient experiment designs), result communication  (e.g., help to interpret ambiguous findings), and research explorations (e.g., formulate directions for further investigations).

In such a fully integrated and engaged partnership, we can bring to the enterprise of scientific justification a valuable \textit{disinterested} --- in its original sense as in ``giving a disinterested advice''--- perspective. Disinterested is not \textit{uninterested} (though unfortunately the former is often misused to mean the latter). I just argued that statisticians should feel the same ownership and accountability as our scientific collaborators, and few can sustain such a feeling without having some genuine interests in the problems themselves.  A disinterested perspective builds upon insights and lessons from similar issues but manifested in different settings, and it provides us with the freedom to explore and scrutinize without being overly attached to the specifics of the problem at hand. An appropriate amount of detachment helps us to reduce the chance of getting onto one of many ``over-paths":  over-fitting, over-interpreting, over-adjusting, etc.   

A good analogy here is how teachers and parents work together effectively in child education. A teacher typically does not know a child nearly as comprehensively as the child's parents, who invest in their child, emotionally and otherwise, far more than a teacher possibly can even if the teacher wants to. Nevertheless, a teacher sometimes can be more effective in diagnosing, and remedying, a child's learning problems than the parents can precisely because of the teacher's freedom to place the child at a suitable distance, and in the broader context of similar problems of other children. This helps the teacher to gain a clearer and larger picture, which in turn can help to form and implement a better remedial strategy.

As a small illustration, if I considered my job as merely being a passive translator and analyst, then I would not bother to ask for the justification of the model 
$C_{ij}=A_iF_j$ in the concordance problem.  Here $A_i$ is the effective area of the $i$th instrument, which roughly speaking is the (equivalent) geometric area  of an instrument (e.g., telescope) that is effective for collecting photons; and $F_j$ is the flux of the $j$th source (e.g., a remnant of a supernova), which, in layman's terms, is a measure of brightness.
I'd then go ahead to fit the Poisson model to data, which are photon counts $\{c_{ij}\}$ with mean $C_{ij}=A_iF_j$, and make inference about $A_i$ and $F_j$ using either the likelihood approach or Bayesian approach, both of which are \textit{statistically principled} (Section~\ref{sec:stat}). This, however, does not guarantee that my results are \textit{scientifically justified}, since that depends on whether they can be discredited by known scientific knowledge.

Having worked on other astrophysical and geophysical projects \citep[e.g.,][]{weatherhead1998factors}, I possess a reasonable amount of prior knowledge that the so-called ``hard physics models'' often have soft spots. In this case, the multiplicative model amounts to assuming that there is no interaction effect, on the log-scale, between instruments and sources. This assumption reduces the number of unknown mean parameters from $N\times M$ to $N+M$. Whereas this is a crucial reduction that makes the problem much more manageable, this very fact also suggests that our conclusions would be rather sensitive to its  validity.  A double check is therefore in order.

Lo and behold, this assumption turns out to be a convenient approximation, because the true mean intensity should be written as $C_{ij}=T_{ij}A_iF_j$, which of course is a tautology without restriction on $T_{ij}$. Besides the obvious and known factors such as the exposure time, $T_{ij}$ also contains ad hoc factors used, for example, to adjust for the ``pileup'' effect, which refers to the situation where several photons hit the detector at the same time,  invalidating  the aforementioned Poisson model. 
The astrophysicists are well aware of this issue, and hence any analysis without addressing it, when it is known to affect the data, would not be considered by them as scientifically justified.  In general, few of our scientific collaborators would knowingly ignore any major defects in the data, but they may not always feel the necessity to disclose such issues or their methods for corrections (the so-called preprocessing), especially in the early stages of collaborations. This is where a statistician's proactive mindset is much needed, because preprocessing is a very messy business with costly consequences \citep[e.g.,][]{blocker2013potential}.  

Using ad hoc adjustment factors, such as $T_{ij}$, to correct for defects in the data (or model) is a popular practice in scientific studies,  because of their simplicity in implementation and interpretation.  Simplicity is of great importance in practice, but these ad hoc adjustment factors often come with uncertainties and errors in themselves, which are typically ignored before we statisticians get involved, for a practical reason. To properly take into account such preprocessing uncertainties and inaccuracies typically requires statistical sophistication that is no less than implementing  statistically principled methods---instead of relying on ad hoc adjustments---for handling data defects in the first place. Hence, many non-statisticians feel uncomfortable executing such methods on their own, or may not see the need to do so.  They understand the potential consequences of not distinguishing between estimated and known quantities, but they may not have seen enough examples in a variety of settings to fully appreciate the extent of the damage that can be caused by ignoring the uncertainty and errors in many seemingly intuitive ad hoc adjustments. Being more proactive means that we should always ask our collaborators about the soft spots in their models and the preprocessing of their data, before getting busy with translating what they told us. Although my experiences are of sample size of one, I yet need to encounter a case where my inquiries (plural intended) did not result in ``Well, Xiao-Li, since you really want to know  ...'' or alike. 

Before I report the principled statistical methods used for the concordance project and how the understanding of preprocessing impacted my modeling strategy, I want to stress that the term ``scientifically justified" is used here to contrast not merely ``scientifically unjustified,'' but also ``scientifically ideal". Since scientific justifications are evidence-based (permitting both past and current evidences), they are inevitably limited by the present theory, understanding or consensus. A scientifically verifiable improvement constitutes a scientific justification, even if the improvement itself is known to be an approximation at the best or even fundamentally flawed in some aspects.  Hence ``scientifically justified'' is not a political or grant-seeking rhetorical.
Rather, it is about not letting either the bad or the perfect be the enemy of the good, so that we can advance sciences in effective and efficient  ways. 

\section{Statistically principled,  not just verified}
\label{sec:stat}

An astute reader may have noticed a confusion between the notation $C_{ij}$ in Section~\ref{sec:cons}, where it represents an observation, and in Section~\ref{sec:stat}, where it represents its expectation, with the observation denoted by $c_{ij}$. 
In statistics, this confusion would qualify to be a dis-qualifier of deserving the title of ``statistician''. But outside of statistics, the distinction between estimates and estimands is not always made or understood, and indeed sometimes it is dismissed as an academic complication created by statisticians. I once overheard an engineer becoming  rather frustrated with a statistician: ``What do you mean that there is an underlying real mean?  What I observe is real and that's all I have and care about!"  

This confusion/frustration reflects a profound defect in our current education system worldwide: starting from kindergarten, and for at least a decade after that, we infuse almost all developing minds with deterministic thinking and operations. Only after student brains are fully developed do our curricula start to include courses with stochastic thinking as their foci. But even then, uncertainties are perceived, and indeed often taught, as annoying ``noise'' rather than being the essences of nature, human, and everything in between. Rarely do we emphasize that uncertainties and information, a.k.a., noises and signals, are the two sides of the same coin: variations.  At the same time, being comfortable, and ideally fluent, in probabilistic language is essential for conducting principled statistical analysis. Without understanding variations, how could any deterministic mind comprehend the fact that the best regression line for predicting $Y$ from $X$ is $Y=\alpha+\beta X$, yet predicting $X$ from $Y$ by reversing this relationship, that is, $X=(Y-\alpha)/\beta$, is a telling sign of being statistically challenged?   

Undoubtedly it will take decades if not centuries for our education system to embrace a ``stochastic first, deterministic second" curricula, or the \textit{kidstogram} approach \citep{meng2018ASA}.  Meanwhile, we statisticians can all take a more active role in communicating statistical principles whenever opportunities arise, including during research collaborations.  Realizing how confused I was by the notation of $C_{ij}$ myself in the initial communications, I started my IACHEC presentation, which by then had became a research proposal instead of a tutorial, with an emphasis on distinguishing between deterministic estimands (e.g, $C_{ij}, A_i, F_j$) and random estimators/data (e.g., $c_{ij}, a_i, f_j$); and the aforementioned ``regression paradox" was used to illustrate why this distinction is critical.  Seeing how the deterministic relationship $C_{ij}=A_iF_j$ had led to the ratio estimator $f_j=c_{ij}/a_i$ as conveyed in the first email cited above (where the notation $C_{ij}/A_i$ was used), I further emphasized that the deterministic relationship among the parameters such as $\log C_{ij}=\log A_i + \log F_j$ does not imply an analogous regression model at the data level. That is, we cannot write $\log c_{ij} = \log a_i+ \log f_j +\epsilon_{ij}$, \textit{and} assume that $\epsilon_{ij}$ has mean zero and is independent of $\{a_i, f_j\}$. Whereas this point is rather obvious to statisticians, it may not be so to those who are uncomfortable with thinking about stochastic relationships or conditional independence. 

While taking logarithms to turn multiplication into addition is a standard strategy, in this case there is a deeper statistical principle to support considering the log transformation.  If the aforementioned multiplicative factor $T_{ij}$, as in $C_{ij}=T_{ij}A_iF_j$, can be regarded as known because its variability has no appreciable impact on the concordance results, then we can simply fit the aforementioned Poisson model. However, astrophysicists do not have such confidence, nor do they assure me that the adjustment $T_{ij}$ as they provide has captured all the important interactions between instrument $i$ and source $j$, beyond those that are captured by $A_i F_j$.  The Poisson model is then too restrictive, and indeed the preprocessed data in the form of $c_{ij}'=c_{ij}/\hat T_{ij}$ are no longer even counts.  These considerations led to the adoption of a log-normal model, especially because the counts in the concordance problems tends to be large.  The separation of the mean and variance parameters in the log normal model is better suited for adhering to the principle of incorporating as many as possible known and potentially influential uncertainties in a statistical model. The normality has the added benefit of computational efficiency (Section~\ref{sec:comp}). Using normal (albeit not log-normal) to approximate Poisson counts is  a common practice in the astrophysical community, so the log-normal approach did not appear to be too exotic to my astrophysicist collaborators. 

The ``half-variance correction" (HVC) proposed in my IACHEC presentation, however, was a bit exotic initially to the audience.  Again, it was driven by a statistical modeling principle: model flexibilities for capturing uncertainties or imperfection should not come at the expense of justified scientific constraints. For the concordance project, we need to ensure that the log-normal model will respect the multiplicative nature of the mean on the original scale. This is the type of modeling considerations that we statisticians are trained to provide. For example, we know well that if $\log y \sim N(\mu, \sigma^2)$, then the mean of $y$ is not $e^{\mu}$, but rather $e^{\mu+0.5\sigma^2}$. Consequently, if we want to model $\log c$ as $N(\mu, \sigma^2)$ but still to ensure $E(c)=TAF$, then we must set $\mu = \log T+ \log A + \log F - 0.5\sigma^2$, hence the HVC.  

The dramatically increased amount and complexity of data in our society has encouraged many to develop and apply a host of methods that are justified solely through simulation studies and empirical validations.  Whereas simulations and empirical studies are indispensable, their scope is inherently limited, and as such they come with prize and price. A principled method may also need empirical validation, and it may not outperform an ad hoc one in any particular application, especially when the evaluative metric is of ``instant gratification'' nature, such as short-term predictions. But just as we tend to trust a seasoned surgeon with our lives more than a rookie one, even if the latter has a higher overall success rate (due to Simpson's paradox), we have a much higher chance of avoiding irreplicable research findings when we push ourselves as hard as we can to follow principled methods, i.e., those come with theoretical understanding of \textit{why and when}, in addition to \textit{how}, they work. 

Being principled is by no means a decree to straitjacket ourselves with traditions and established rules.  To the contrary, major methodological advances occur when we develop sound theory and principles to explain and improve seemingly ad hoc methods that appear to work well from empirical evidence. Indeed, it is virtually impossible to find a time-honored statistical method with unknown rationale or scope of its applicability and limitations. The current quest for a deep understanding of deep learning is a reflection of our time-honored desire to be principled.    

The rewards, and hence the joy, for being statistically principled are not just better science, but also richer statistical methods and insight.  As a small example, HVC forces the mean of  the log normal to depend on (but not determine) its variance. This leads to intriguing new shrinkage phenomena. For instance, under the hierarchical log-normal model with HVC, not only the maximum likelihood estimators of the means are of the shrinkage forms, but also the variance estimator itself. However, the latter is in unfamiliar  self-weighted shrinkage form, $\widehat\sigma^2= R S^2_{y}$, where $R=\frac{2}{1+\sqrt{1+S^2_{y}}} \le 1$
and $S_y^2$ is in the form of the usual regression residual variance in the absence of HVC \citep{yang2018lognormal}. This provides a new statistical insight on how such model works: there is a self-balancing act. A small $S^2_y$ indicates little need of HVC, and hence little modification to $S^2_{y}$ itself is called for:  $R\uparrow 1$ as $S^2_{y}\downarrow 0$. A large $S^2_{y}$ indicates the need of HVC, which then should reduce the residual variance. However, this in turn would reduce the estimator of $\sigma^2$, which then should trigger a smaller HVC, which then could lead to too large a residual variance if we over-correct. Whereas we could engage ourselves in this indefinite and heuristic balancing act,  the precise form of $R$ here tells us exactly what the optimal 
shrinkage factor should be. Such non-linear and self-weighting factors would be hard to construct without following principled methods such as MLE.  Indeed, even if $R$ were conjured up by an ingenious mind, a principled justification would still be needed in order to ensure that it is optimal, or even just to convince ourselves (and others) why such an exotic looking factor  makes sense at all.  

\section{Computationally efficient, not simply reproducible}\label{sec:comp}

The growing demand for handling large and complex data sets has increased sharply the awareness in our statistical community that our students need more training in computer science, from data processing to data curation and to algorithm design and implementations. Our traditional training is simply inadequate for dealing with the volume, variety and velocity of the data as we now see every day, or rather every second. I, for one, can testify to the inadequacy of such training at least for my generation.  Although  1/3 of my thesis was on the EM-type of algorithms, and roughly 1/3 of my research has been on statistical computing, my ability for data processing or algorithm implementation has regressed to the level that I simply do not even try, because that would waste everyone's time, not just mine.  Having never taken a course in computer science, I struggled through by learning from fellow students and by doing things in ad hoc ways. A vivid example was when I was taking my Ph.D. qualifying exam, I had to use three different languages/packages (Minitab, Gauss, and Fortran 77), because I knew how to use each only for some specific tasks. Thank God that my professors were in the same camp, in the sense that all they cared about was the correctness of my output, not about how I computed them.  

The long-term consequence, however, is not of the thank-God nature. If not for my many wonderful students and co-authors, I'd not be able to contribute (directly) to any project involving a decent amount of data. Indeed, without two very able Ph.D. students who joined the concordance project, the half variance correction would remain as a half baked idea.  Having two students working on the same project is a luxury that not everyone can afford, but for those who can, I highly recommend it. This is not merely about having more brains and hands, but about independent verifications using different algorithms/packages/programming languages. With such independent verifications, the assurance of computational reproducibility, as a minimal requirement for scientific replicability, is greatly increased.   

The existence of multiple algorithms for the same problem, which is almost always the case for fitting statistical models, opens the door for better efficiency. Traditional considerations for computational efficiency in statistics include convergence rates and number of iterations, CPU time, storage space, and occasionally human time (e.g., for programming); there are also considerations for computational stabilities \citep[e.g.,][]{van2010cross}. The arrival of Big Data, however the term is defined, has reconfirmed the values of these considerations. But more pressingly, it brought additional ones, such as scalability, transportability, and parallelizability, terms that I can write about only with wiki-depth.   
And I doubt I am a minority among my contemporaries in this regard.  To avoid such inadequacy for the future generations, those of us who have the responsibilities and opportunities to design curriculum, should encourage, or even require, our students to take courses from our computer science colleagues.

Meanwhile, we can still do our best to ensure the computation for our collaborative projects is carried out as efficiently as possible, whenever we have to do it ourselves. The process of searching for the most efficient algorithms itself can serve multiple purposes. For example, computational difficulties often reflect modeling challenges (such as near-nonidentifiability), known as a \textit{The Folk Theorem of Statistical Computing} {({\tt http://andrewgelman.com/2008/05/13/the\_folk\_theore})}. Recognizing this tendency  would lead naturally to increased effort to improve the statistical model itself. Efficient computation also helps future investigations and investigators of similar problems. Furthermore, it provides very valuable training for our students, in terms of establishing healthy research mindsets (e.g., seeking the best possible method, instead of the least-publishable unit), as well as building richer and more rigorous research processes. It can also provide unexpected but potentially fruitful research topics. 

For the aforementioned log-normal model, there are at least four algorithms to fit it Bayesianly. Three of them were implemented in \cite{yang2018lognormal}: a vanilla Gibbs sampler, an improved Block Gibbs, and a Hamiltonian Monte Carlo (HMC) algorithm. As it is well-know, Gibbs samplers tend to save human time, being
easy to implement with little tuning, but at the expense of running time.   HMC is the opposite, requiring careful tuning, but runs fast once tuned properly. Both phenomena occurred for the concordance project, but the nearly identical output from the Block Gibbs and HMC endowed us with greater confidence in their validity than would be possible if only one of them had been implemented.

Most intriguingly, the fourth algorithm was discovered accidentally during our proof of the propriety of the posterior under an improper prior. Such proof should always be required as a part of our overall emphasis on being statistically principled. During the proof, we discovered a simple 
bound for implementing standard rejection sampling to sample from the marginal posterior of the variance parameters. Because the conditional posterior of the mean parameters given the variance parameters is multivariate Gaussian, we can then bypass MCMC entirely, and produce the ideal i.i.d. draws. This finding may have implications to other problems involving multiplicative signals but additive noises. 

Indeed, CHASC has a history of turning implementation tricks from a project into a general computational strategy. For example, \cite{yu2011center} documented how a seemingly simple idea for fitting a Cox process model (for studying a neutron/quark star) led to a general interweaving strategy for improving MCMC, sometimes dramatically. A very successful case, the ancillary-sufficient interweaving strategy (ASIS), was identified because of the principled statistical thinking via classical concepts such as ancillarity and sufficiency, as well as the ``regression paradox" mentioned in Section~\ref{sec:stat}  
\citep{xu2013thank}. ASIS has helped (astro)physicists to study sources such as black holes and quasars (e.g., \cite{kelly2013active,krawczyk2015mining}). More encouragingly, ASIS enabled a modeling fitting in finance that was otherwise infeasible, as reported by \citet{kastner2014ancillarity} and \citet{kastner2016dealing}. 

Most recently, an effort to improve MCMC for fitting a latent Ornstein-Uhlenbeck process model for accessing time delays led to a new kind of Metropolis algorithm that is more effective for multi-modal distributions: the Repelling-Attracting Metropolis (RAM) algorithm, and its utility and potential has been explored beyond astrophysical applications, as detailed in \cite{tak2018repelling}. Seeing scientific applications turn into methodological advances is always a joy, at least for those of us who care about advancing the science of data, concurrent with advancing science through data. 

\section{Principled corner cutting, not cutting principles} 

Conducting highly principled data science cannot be rushed. Hence it is not for those who seek fast publications --- a mantra for such researchers is always that ``hard to publish, but impossible to unpublish." It however helps to advance science at a faster pace. Unreplicable results not only waste resources and erode the public trust in science, but most damagingly they can seriously mislead and even derail scientific advancement itself. Research speed can be gained by principled corner cutting \citep{meng2018ASA}, but not by cutting principles. That is, we seek compromises and shortcuts only after we understand their consequences, and know how to put the corners back, and in what order, when more time or resources become available. As we statisticians become main players on the central scientific stages, it is essential for us to hold a very high bar for ourselves, just as great scientists always do. The tribute Vinay and Aneta wrote for Alanna reminds us how great scientists devote their lives to being principled, which I quote below both to pay my respects to Alanna and to thank my CHASC collaborators for their many years of inspirations and patience with me, as I aim to grow from a consultant to a co-investigator. 

``{\sl Alanna was a gamma-ray astrophysicist, and was at the forefront of Astrostatistics and Astroinformatics.  She was a strong advocate for principled analyses, and was a pioneer in the development and application of Bayesian methods to Astrophysical problems. She was a pivotal and founding member of our international Astrostatistics collaboration. She was instrumental in organizing dozens of workshops to bring the awareness and knowledge of new ideas to the community.
She was our beloved friend and admired colleague. We will greatly
miss her presence, her wisdom, and her determination.}''

\baselineskip=11pt
\bibliographystyle{apalike}
 
\bibliography{calibrationbib}

\end{document}